\documentclass[aps,prd,preprintnumbers,showpacs]{revtex4}
\usepackage[dvips]{graphicx}
\begin{document}

%
%

\eprint{Nisho-1-2015}
\title{Production of Semi Quark Gluon Monopole Plasma by Glasma Decay}
\author{Aiichi Iwazaki}
\affiliation{International Economics and Politics, Nishogakusha University,\\ 
6-16 3-bantyo Chiyoda Tokyo 102-8336, Japan.}   
\date{Nov. 8, 2015}
\begin{abstract}
Using  the standard Lagrangian of  gluons and a model of dual superconductor for magnetic monopoles,
we calculate the number densities of the gluons and monopoles 
produced by the decay of background 
color electric $E$ and magnetic $B$ fields ( glasma ).
We find that gluons are dominant decay products 
when the initial values of the gauge fields are large
such that $gE=gB>(0.5\mbox{GeV})^2$,
while they are suppressed and 
monopoles are dominant decay products
when the initial values are small such that $gE=gB<(0.4\mbox{GeV})^2$.
The feature of the gluon dominance at large $gE=gB$ and the monopole dominance at small $gE=gB$ 
is similar to the one of thermalized quark gluon monopole plasmas proposed recently,
if we identify $\sqrt{gE}=\sqrt{gB}$ as temperatures of the plasmas.
The identification is suggested by the fact that the energy densities of the gluons and monopoles 
are proportional to the initial values $(gB)^2=(gE)^2$, while the energy densities of the plasmas are proportional to $T^4$.  
The feature of the gluon dominance in the glasmas with large saturation momenta has been derived in classical statistical field theories, while
the feature of the monopole dominance has not yet derived. 
Although the model of the monopoles is phenomenological, our analysis suggests that
the monopoles play important roles in the decay of the glasmas with small saturation momenta, to which
classical statistical field theories are not applicable. 
\end{abstract}
\hspace*{0.3cm}
\hspace*{1cm}

\maketitle


\section{introduction}

Quark gluon plasmas ( QGPs ) have been produced by high energy heavy ion collisions.
They have been extensively explored and shown to be thermalized\cite{hirano} in a very short period
$<1$fm/c after the collisions. The plasmas are expected to approach to the ideal
gas at high temperatures $T \gg 1$GeV. In such a temperature the plasmas
are composed of weakly coupled quarks and gluons. On the other hand,
the plasmas are composed of strongly coupled quarks and gluons\cite{splasma} at low temperatures, e.g.
$T<0.7$GeV. With further decrease of the
temperatures, a phase transition takes place at a critical temperature $T_c\sim 0.16$GeV and 
the quarks and gluons are confined in hadrons.

At high temperatures, the weakly coupled quarks and gluons
are quasi-particles in the plasmas.
On the other hand,  it is not clearly understood what are
quasi-particles in the plasmas of the strongly coupled quarks and gluons
at low temperatures,
for example, 
at the temperature $T=(1\sim 2)T_c$ where
coupling strength is given by
$\alpha_s=g^2/4\pi\sim 1$ with
gauge coupling constant $g$. 
But, the number $f_{\rm eff}$ of quasi-particles 
defined such that $f_{\rm eff}\equiv 30\epsilon/(\pi^2T^4)$ with energy density $\epsilon$ has been shown in the lattice gauge theories\cite{lattice}
to be rapidly suppressed 
in such strong coupled plasmas as the temperatures approach to the critical temperature $T_c$.

It has recently been proposed\cite{gyu} that quasi-particles of the strong coupled QCD plasmas
are magnetic monopoles\cite{monopole} in addition to quarks and gluons. 
According to the model \cite{gyu},
quarks and gluons are dominant components in the plasmas at high temperature $T > 3T_c$, while the monopoles are 
dominant components at low temperature $T=(1\sim 2)T_c$. In other words, the effective dynamical degrees of freedom of  
quarks and gluons are suppressed in the low temperature $T\simeq (1\sim 2)T_c$,
and instead the monopoles become dominant. 
It was pointed out\cite{mono} before the proposal\cite{gyu} 
that the monopoles play important roles in strongly coupled QGPs with temperatures near $T_c$.
For example, they play a role of making small
shear viscosity of the plasmas\cite{mono}. Furthermore,
the monopoles play the role of quark confinement\cite{t,ie} at the temperatures $T<T_c$.
The dominance of the monopoles at low temperature
$T\simeq (1\sim 2)T_c$ and the dominance of the quarks and gluons at high temperature $T > 3T_c$ is a 
characteristic feature of the model of the thermalized quark gluon monopole plasmas ( QGMPs ). 

\vspace{0.1cm}
In this paper we discuss prethermalized states of monopoles and gluons. 
They are the states produced by 
the decay of homogeneous color electric $E$ and magnetic $B$ fields. 
The presence of such classical gauge fields ( glasmas ) produced by 
the high energy heavy ion collisions has been discussed in a model of color glass condensate\cite{cgc}. 
Although these monopoles and gluons interact with each other and would be thermalized after their production,
the states we discuss in the paper are prethermalized states of the non-interacting gluons and monopoles. 
We do not address how the states are thermalized, but address which ones are dominant decay products of 
the gauge fields, gluons or monopoles.

\vspace{0.1cm}
The field strengths of the glasmas depend on saturation momentum $Q_s$ of the color glass condensate.
The strong gauge fields with large $Q_s$ have small gauge couplings  $\alpha_s\ll 1$, 
while the weak gauge fields with small $Q_s$, but still larger than $\Lambda_{QCD}$
have large gauge couplings $\alpha_s\sim 1$.  
There are some reliable methods with which 
the decays of the strong gauge fields can be analyzed. They are classical statistical field theories\cite{csft1,csft2,iwazaki},
Schwinger mechanism\cite{sch,tanji} or classical numerical simulations\cite{ven,berges,kunihiro,fuku} in gauge theories of quarks and gluons.
But there are no reliable methods with which the decays of the weak gauge fields can be analyzed
since $\alpha_s$ is large. 
We need to use some non-perturbative methods or models of strongly coupled quarks and gluons.
Here we use a model of dual superconductors\cite{dual,koma} in QCD as an effective field theoretical model of the monopoles;
they are assumed to be quasi-particles of strongly coupled gluons.

The phenomenological model of the monopoles has been used for the analysis of quark confinement in QCD vacuum.
It is natural to apply it to the analysis of the glasma decay which leads to strongly coupled QGPs, in which
perturbative analysis is not valid. The non-perturbative analysis based on the phenomenological
model would be valid for example in a range $1>\alpha_s>0.5$ ( or $T\simeq (1\sim 2)T_c$ ) 
where the glasmas still hold the coherence as classical fields and the monopoles do not strongly couple with each others.
We note that the gluon occupation number of the glasmas is of the order of $\alpha_s^{-1}$ and the magnetic couplings
of the monopoles are also given by $\alpha_s^{-1}$.
For the analysis of the glasma decays, 
we use both of the model and the gauge theories of gluons in the whole range of the gauge couplings
$1>\alpha_s>0$. 
It turns out that the glasmas mainly decay into monopoles when the gauge couplings are strong such as $1>\alpha_s>0.5$,
while they mainly decay into gluons when  they are weak such as $0.3>\alpha_s>0$.

The prethermalized states of the decay products have a similar feature to that in 
the QGMPs mentioned above. 
Namely, the gluons are dominant decay products of
the strong gauge fields, while the monopoles are dominant ones of
the weak gauge fields.
Thus, the temperatures of the thermalized plasmas produced by the strong gauge fields
is high, while the temperatures of the thermalized plasmas produced by the weak gauge fields
is low. 
Thus, if we identify the initial values of $\sqrt{gB}$ or $\sqrt{gE}$ as a temperature, 
the dominant components of the prethermalized plasmas
are similar to the ones in the QGMPs.  
The identification $\sqrt{gB}$ or $\sqrt{gE}\sim T$ is suggested by the fact that the energy densities of the prethermalized gluons and monopoles 
are given by the initial values $(E^2+B^2)/2$ of the gauge fields,
while those of the QGMPs are roughly given by $T^4$. 
There is a duality such that the gluons play dominant roles in the glasmas with large $Q_s$ or QGP plasmas with high temperatures, while
the monopoles do dominant roles in the glasmas with small $Q_s$ ( but still larger than $\Lambda_{QCD}$ ) 
or QGP plasmas with low temperatures ( but still larger than $T_c$ ). 

\vspace{0.1cm}
We assume in the present paper that the color glass condensates are still present even when $Q_s$ is small just as $Q_s= (1\sim 2)\Lambda_{QCD}$
and that they are described by classical color gauge fields $E$ and $B$ since the gluon occupation number is given by $\alpha_s(Q_s)^{-1}\simeq 1\sim 2$ . 
We call such glasmas strongly coupled glasmas, while we call the glasmas with large $Q_s$ weakly coupled glasmas.
In the next section, we describe how the monopoles play roles in the decay of the strongly coupled glasmas.
In the section \ref{3}, 
we describe the applicability of our model by the use of which the productions of gluons and monopoles are discussed.
Actual models of gluons and monopoles are presented in section \ref{4}.
The evolution equations of the number densities of gluons and monopoles are presented in section \ref{5}.
Our results are shown in the section \ref{6}. In the final section we present our discussions and conclusion.

\section{magnetic monopoles}

We explain the role of the color magnetic monopoles in the decay of the glasmas.
Their decay has been mainly discussed using the classical statistical field theory\cite{csft1,csft2,iwazaki}, 
Schwinger mechanism\cite{sch,tanji} or classical numerical simulations\cite{ven,berges,kunihiro,fuku}. 
Although the classical statistical field theory is well controlled method, it is
only applicable to the very weakly coupled glasmas, 
that is, glasmas with sufficiently large saturation momenta $Q_s$ for the gauge coupling to be extremely small $g(Q_s)\ll 1$. 
However, the theories are not applicable for the glasmas with realistic small gauge couplings $g(Q_s) \sim O(1)$\cite{gelis}.
On the other hand, the Schwinger mechanism is applicable even for the moderately strongly coupled glasmas. 
The pair creation of gluons arise according to the mechanism, which makes the color electric fields decrease. 
But the magnetic fields of the glasmas hardly decay in the mechanism.
This is because the pair creations of gluons do not make the magnetic fields decrease.
Similarly,
the numerical simulations using classical equations of motion are only applicable for sufficiently strong gauge fields
for gluons to keep the coherence. But with the expansion of the glasmas, the coherence is not kept
since the gluons become dilute. Furthermore, the classical treatments including the classical statistical studies does not make clear what are quasi particles 
after the decay of the gauge fields.
In this way,
it is not efficient to apply these methods to the analysis of the decay of the strongly coupled glamsas
with saturation momenta such as  $\alpha_g(Q_s)= 1/2\sim 1$.
 
Obviously,
the magnetic monopoles make the magnetic fields efficiently decay in monopole plasmas. 
They also play the role of confining quarks and gluons\cite{t,ie} in the strongly coupled QCD vacuum.
The monopoles are well defined objects in QCD when the gauge couplings are large, since the magnetic charge is
so small that their mutual interactions are small. 
It is expected that the monopoles play important roles in the glasmas with small $Q_s$ or QGPs with temperatures near $T_c$.
Actually, it was pointed out\cite{mono} that the monopoles play important roles in the strongly coupled QGPs with the low temperatures
as well as in QCD vacuum.
In particular, it has recently been discussed\cite{gyu} in the realistic analysis of high energy heavy ion collisions
that they are present even at $T>T_c$ and
play significant roles in the QGMPs. In the discussions they are treated simply as
point particles with magnetic charges $g_m$ satisfying the Dirac quantization condition $g_m=4\pi/g$.
But, their production mechanism in heavy ion collisions and their properties ( masses or spins )
in the thermalized states are 
still not well-known. 
Thus, it is reasonable to apply a phenomenological model of the magnetic monopoles to the analysis of the decay of the strongly coupled glasmas
with small $Q_s$. 
It is the model of the dual superconductors. It has been extensively discussed to analyze 
strongly coupled QCD vacuum. 
The model is phenomenological and our production mechanism of the monopoles is rough. But,
our results are consistent with the model of the QGMPs;
the monopoles is strongly suppressed in the states arising from weakly coupled glasmas ( in the QGMPs with high temperatures ), 
while they are dominant in the states arising from the strongly coupled glamsas ( in the QGMPs with low temperatures ).

\vspace{0.1cm}
Here we mention that the classical gauge fields of the glasmas are present
when the gluon occupation number  $\propto \alpha_s(Q_s)^{-1}$ in the color glass condensates is much larger than unity.
That is, the coherence of the gluons is present for $\alpha_s(Q_s) \ll 1$. It is realized for large saturation momentum $Q_s\gg \Lambda_{QCD}$.
In the range, the classical statistical field theories are applicable to the decay of the glasmas, resulting in the gluon production. 
When $Q_s$ becomes smaller, the coherence of the gluons gets worse.
The strongly coupled glasmas we discuss are characterized by large gauge couplings, but we expect that
the coherence of the gluons still holds. We may assume that the classical gauge fields of the glasmas are 
present even for large gauge couplings such as $\alpha_s \lesssim 1$ ( or $Q_s\gtrsim \Lambda_{QCD}$ );
the occupation number is of the order of unity for $\alpha_s \simeq 1$.
Therefore, the phenomenological model of the monopoles, which would be valid for large gauge coupling such as $\alpha_s\gtrsim 1/2$,
 can be applied to the decay of the classical gauge fields of the glasmas.

\section{applicability of Schwinger mechanism}
\label{3}

Our production mechanism of gluons and monopoles is Schwinger mechanism, that is, 
they are generated as pair production\cite{sch,tanji} under the background color electric and magnetic
fields. We assume that the background gauge fields are spatially homogeneous
and are pointed into the identical directions, both in real and color spaces.
The gauge fields decrease with the pair production of the color charged particles.
Furthermore, we assume that the field strength of color electric and magnetic fields are initially
identical; $gE=gB=Q^2$.
We have a parameter $Q$ representing the strength of the gauge fields. We should point out that
the energy densities of the gluons and monopoles produced are given by $Q^4/g^2$,
since the energies of the gauge fields are transformed into the energies of the particles.

As we show below, most of the gluons produced by Schwinger mechanism are the ones
called as Nielsen-Olesen unstable modes\cite{nielsen}.
The modes arise when classical color magnetic fields are present.
Their presence implies instabilities\cite{instability} of the gauge fields and
has been discussed
in several numerical simulations\cite{ven,berges,kunihiro,fuku} using 
inhomogeneous background gauge fields.
The growth rates $\gamma$ of the exponentially growing unstable modes 
$\sim \exp(\gamma t)$ found in the simulations
correspond to $Q$ in the present paper, i.e. $\gamma=Q$.
That is, we describe the instabilities arising under inhomogeneous background gauge fields
as instabilities arising under the
homogeneous background gauge fields\cite{instability}. The field strengths of the homogeneous gauge fields 
are appropriately chosen so as to
give rise to the identical growth rates to the ones obtained in the numerical simulations with the inhomogeneous background gauge fields.
The description using such homogeneous gauge fields 
may be considered as a mean field approximation for
gauge fields with general inhomogeneous configurations.
In general, the parameter $Q$ is much less than real saturation momenta $Q_s$ of glasmas.
The calculation of the gluon production in the Schwinger mechanism is only reliable in the glasmas with large $Q$ such as $\alpha_s(Q)\ll 1$,
since the gluons must weakly interact with each other for our approximation to be valid. 

When we naively apply it to the glasmas with small $Q$ ( $\alpha_s(Q)\sim 1$ ),
 we find that the electric fields decay so slowly that the gluon production
hardly arise. 
But the application is not suitable to the glasmas.
Then we need to see how the electric fields decay after the rapid decay of the magnetic fields. We show in the section \ref{7} that
the energies of the electric fields dissipate in the monopole plasmas without the gluon production. 
Thus, the result of the gluon suppression is valid.

\vspace{0.1cm}
On the other hand,
we assume an effective field theoretical model of monopoles in order to calculate their productions
in the Schwinger mechanism.
The model describes dual-superconductors\cite{dual,koma} in which quark confinement is realized 
with monopole condensations.
We apply it to the analysis of the states in which gluons
strongly couple with each other and the monopoles weakly couple with each other.
The QGMPs with low temperatures as well as
the prethermalized states 
produced by the decay of weakly coupled glasmas with small $Q$ would be such states.
But the model is not applicable to the states with high temperatures $T\gg T_c$ or the weakly coupled glasmas with large $Q$,
since the magnetic charge $g_m=4\pi/g$ is large so that the monopoles strongly couple with each other. 
On the other hand,
it is general consensus that the monopoles do not play
any roles and are absent in weakly coupled QGPs.
Our result is consistent with the QGMPs;
the monopoles is strongly suppressed in the prethermalized states arising from the glasmas with large $Q$, 
while they are dominant in the prethermalized states arising from the glasmas with small $Q$.

When we naively apply the model to the glasmas with large $Q$ ( $\alpha_s(Q)\ll 1$ ), we find that the magnetic fields decay so slowly 
that the monopole production hardly arise.
But the application is not suitable to the glasmas. As we show in the section \ref{7},
the energies of the magnetic fields dissipate in the gluon plasmas without the monopole production.
Thus, the result of the monopole suppression is valid.

\section{models of gluons and monopoles}
\label{4}

First we explain our model.
We consider gluons in SU(2) gauge theory with
the background color electric and magnetic fields given by
$\vec{E}_a=\delta_{a,3}(0,0,E)$ and $\vec{B}_a=\delta_{a,3}(0,0,B)$; $a=1,2,3$.
They are supposed to be spatially homogeneous and
collinear both in the real and color spaces.
The gauge fields are represented 
by the diagonal component of the gauge potential $A_{\mu}\equiv A_{\mu}^{a=3}$.
Under the background fields, the off-diagonal components 
$\Phi_{\mu}\equiv (A_{\mu}^1+iA_{\mu}^2)/\sqrt{2}$ 
perpendicular to $A_{\mu}^3$ behave
as charged vector fields. When we represent SU(2) gauge potentials $A_{\mu}^a$ using 
the variables $A_{\mu}$ and $\Phi_{\mu}$, Lagrangian of SU(2) gauge fields is
written\cite{instability} in the following,

\begin{equation}
\label{L}
L=-\frac{1}{4}F_{\mu,\nu}^2-\frac{1}{2}|D_{\mu}\Phi_{\nu}-D_{\nu}\Phi_{\mu}|^2
-ig(\partial_{\mu}A_{\nu}-\partial_{\nu}A_{\mu})\Phi^{\dagger \mu}\Phi^{\nu}
+\frac{g^2}{4}(\Phi_{\mu}^{\dagger}\Phi_{\nu}-\Phi_{\nu}^{\dagger}\Phi_{\mu})^2,
\end{equation} 
with $F_{\mu,\nu}=\partial_{\mu}A_{\nu}-\partial_{\nu}A_{\mu}$ and 
$D_{\mu}=\partial_{\mu}-igA_{\mu}$.
The gauge field $A_{\mu}$ represents both the background gauge fields $E$ and $B$. 
We find that the fields $\Phi_{\mu}$ represent charged vector fields
with the anomalous magnetic moment described by the term 
$-ig(\partial_{\mu}A_{\nu}-\partial_{\nu}A_{\mu})\Phi^{\dagger \mu}\Phi^{\nu}$. 
They also have standard interactions with the gauge fields $A_{\mu}$ 
through the covariant derivative $D_{\mu}$.
Therefore, it is easy to see that when the background magnetic field 
$B=\partial_1A_2-\partial_2A_1$ is present, 
the gluons represented by the fields $\Phi^{\mu}$
occupy the Landau levels and interact with each other through the term
$\frac{g^2}{4}(\Phi_{\mu}^{\dagger}\Phi_{\nu}-\Phi_{\nu}^{\dagger}\Phi_{\mu})^2$.  
The energies $E_n$ of the gluons with spin parallel to the background magnetic field
are given by $E_n^2=(2n+1)gB-2gB+p_z^2=(2n-1)gB+p_z^2$ with integer $n\ge 0$
where $p_z$ denotes momentum component parallel to $\vec{B}$.
The modes effectively have imaginary mass $i\sqrt{2gB}$, which arises from the 
term of the anomalous magnetic moment.
Thus, we find that the modes with $E_{n=0}=\sqrt{p_z^2-gB}$ are unstable when $p_z^2-gB<0$;
the gauge fields exponentially grow 
such that $\Phi_{NO}\sim \exp(t\sqrt{gB-p_z^2})$.
The modes are called as Nielsen-Olesen unstable modes\cite{nielsen,instability}
and are produced spontaneously under the magnetic field $B$.  
We can see from the exponential growth of the gauge fields $\Phi_{NO}$ that
the modes with smaller $|p_z|$ are produced more abundantly.
The fact indicates that the soft modes of the gluons are dominantly produced
in the early stage of the glasmas decay.
 
When the electric field $E$ is present, the production is accelerated owing to the
Schwinger mechanism. 
As we show later, when the background gauge fields are strong, 
the gluons are dominant decay products of the gauge fields.
It comes from the imaginary mass $i\sqrt{2gB}$ of the gluons.

On the other hand, the energies $E_n$ of the gluons with spins anti-parallel to $\vec{B}$
are given by $E_n^2=(2n+1)gB+2gB+p_z^2=(2n+3)gB+p_z^2$. The modes are stable 
and effectively have mass $\sqrt{2gB}$ arising from the term of the anomalous magnetic moment.
They are produced only when the electric field is present.

Since the production of the gluons and monopoles
eventually makes the background gauge fields $E$ and $B$ vanish,
the effective masses of the gluons vanish. Thus, the gluons becomes massless 
after their production.

\vspace{0.2cm}
Our model of the monopoles\cite{monopole} describing dual superconducting states\cite{ie,dual,koma}
is given by

\begin{equation}
L=|D_{\nu} \phi|^2-\lambda(|\phi|^2-v^2)^2=|D_{\nu} \phi|^2+\mu^2|\phi|^2-\lambda |\phi|^4-
\lambda v^4
\end{equation}
with $\mu^2\equiv 2\lambda v^2$ and $D_{\nu}=\partial_{\nu}-ig_mA_{\nu}^d$,
where the field $\phi$ represents the monopole.
We denote magnetic charge $g_m=4\pi/g$
and dual gauge potential $A_{\nu}^d$. 
We should note that the monopoles have imaginary mass $i\mu$ around the state $\phi=0$. Thus,
the monopole field exponentially grows such that $\phi\sim \exp(t\sqrt{\mu^2-\vec{p}^2}) $.
It implies that the monopoles are spontaneously produced in the state $\phi=0$
even without color magnetic fields $\vec{B}=-\partial_0\vec{A^d}-\vec{\partial}A_0^d$.
Similarly to the case of Nielsen-Olesen unstable modes, the monopoles with soft modes $\vec{p}^2 \ll \mu^2$ 
are dominantly produced and
condense to make a confining vacuum; $\langle \phi \rangle
=\mu/\sqrt{2\lambda}$. 

The state $\phi=0$ arises 
immediately just after the high energy heavy ion collisions. According to a model of color glass condensate,   
only longitudinal color electric and magnetic fields are generated after the collisions. It implies that
there is no overall magnetic charges $\langle\int d\vec{S}\cdot \vec{B}\rangle=0$.
Thus we may suppose that the state $\phi=0$ is initially realized in the glasmas. 
The spontaneous production of the monopoles begins just after the collisions.
When the magnetic field is present, the production of the monopoles is accelerated
owing to the Schwinger mechanism. Furthermore,
when color electric field $E$ is present, the monopoles occupy Landau levels specified
by integer $n\ge 0$. Their energies $E_n^m$ are given by $(E_n^m)^2=(2n+1)g_mE+p_z^2-\mu^2$.
Thus, when the background electric field $g_mE$ is smaller than $\mu^2$,
the monopoles in the lowest Landau level ( $n=0$ ) are spontaneously produced.
On the other hand the spontaneous production does not arise when 
the electric fields are strong enough such as $g_mE>\mu^2$.

We show below with the explicit use of the parameter
$\mu=0.5$GeV that the large amount of the monopoles are dominantly 
produced by the weak gauge fields with $\sqrt{gE}=Q<0.4$GeV$<\mu$,
while the production of the monopoles is suppressed 
for the strong gauge fields with $\sqrt{gE}>\mu=0.5$GeV since the spontaneous production
does not arise.
We find that the values of the imaginary mass $\mu$  
control the critical field strength $gB_c=gE_c=Q_c^2$ beyond which the monopole production is suppressed.

Furthermore, we show that
each of the monopoles abundantly produced 
has small kinetic energies $< 10$MeV for very weak gauge fields such as $\sqrt{gE}<0.3$GeV.
The weaker gauge fields induce the production of much more abundant monopoles
with smaller kinetic energies. 
That fact leads to large collision cross sections $S=\pi l^2$ between monopoles since
the distance $l$ is roughly given by solving the equation such as 
the potential energies $g_m^2/l$ equal to the small kinetic energies of the monopoles; $S\propto 1/(\rm kinetic\,\, energy )^2$.
Thus, after the decay of the magnetic fields with small $Q$, the electric fields would rapidly decay in the monopole plasmas because  
magnetic resistances ( $\propto S$ ) are large for small kinetic energies of the monopoles.

\section{evolution of number densities of gluons and monopoles}
\label{5}

We now proceed to show how the number densities of the gluons and monopoles evolve with time. 
We first note that
the color charged particles are accelerated by color electric or magnetic fields.
Thus, the energies of these gauge fields decrease.
When the number density $n_g$ ( $n_m$ ) of the gluons ( monopoles ) is given,
the energies of the charged particles increasing with their acceleration in a period $dt$
are given such that

\begin{equation}
\label{2}
dt\times n_g\times gE = -d\Bigl(\frac{E^2}{2}\Bigr) \quad \mbox{and} \quad
dt\times n_m\times g_mB = -d\Bigl(\frac{B^2}{2}\Bigr).
\end{equation}
( We have neglected a term associated with polarization current\cite{tanji}, 
which comes from quantum production of the particles.
The term has been shown much smaller 
than the terms in eq(\ref{2}) associated with conduction currents\cite{tanji}. )
The equations govern 
the evolution of the electric $E$ and magnetic $B$ fields as well as the number densities $n_g$ and $n_m$.
In order to solve the equations we need to know the number densities as the functions of
the gauge fields; $n_g(gE,gB)$ and $n_m(gE,gB)$. 
The number densities of the charged particles produced by the Schwinger mechanism,
have been obtained numerically in the references\cite{tanji,ita}, in which
approximate formulas have also been given.
We use the formulas given in the references.

\vspace{0.1cm}
Before giving the number densities of the gluons and monopoles
under the background gauge fields, we notice that the number density of
a charged scalar field with mass $m$ and charge $g$ produced 
by Schwinger mechanism under the electric field $E$
has been given\cite{tanji} by

\begin{equation}
\label{b}
n(t)=\int\frac{d^3p}{(2\pi)^3}\exp(-\frac{\pi(m^2+p_x^2+p_y^2)}{gE})=
\int \frac{gE\,dp_z}{(2\pi)^3}\exp(-\frac{\pi m^2}{gE})\simeq \frac{(gE)^2t}{(2\pi)^3}\exp(-\frac{\pi m^2}{gE}),
\end{equation}
where we have taken into account the allowed range $gEt>p_z>0$ of the momentum $p_z$ of the produced particles
after the electric field is switched on at $t=0$. That is, the production rate of the particles is
proportional to $\exp(-\pi(m^2+p_x^2+p_y^2)/gE)$.
The formula has been given for the particles with their momentum $p_z$ much larger than $m$.  
Furthermore, it is valid only for the 
electric field $E$ constant with time $t$. Hereafter, we assume the formula even for 
$E$ varying with time $t$, as long as the variation is smooth.
As we show below, the gauge fields smoothly decay up to a certain time $t_c$, but decay rapidly
after $t_c$. Thus, we use the formula until the rapid decay starts.
We evaluate the number density $n_g$ and $n_m$ of the particles produced by the decay of the gauge fields at the time $t_c$.  

When we impose a magnetic field $B$ in addition to the electric field $E$,
the number density of the particles with energies $E_n=(2n+1)gB+p_z^2+m^2$ is given by

\begin{equation}
\label{n}
n(t)=\frac{gEt\times gB}{(2\pi)^2}\frac{\exp\bigl(-\frac{\pi(m^2+gB)}{gE}\bigr)}
{1-\exp(-\frac{2\pi gB)}{gE})},
\end{equation} 
where the factor $gB/(2\pi)$ comes from the degeneracy of a Landau level and
the factor $(1-\exp(-\frac{2\pi gB)}{gE}))^{-1}$ comes from the summation, 
$\sum_{n=0}^{n=\infty} \exp(-2n\pi gB/gE)$,
where $n$ denotes the Landau levels.
We should note that the term $m^2+gB$ in eq(\ref{n}) 
corresponds to the term $m^2+p_x^2+p_y^2$ in eq(\ref{b}). That is, 
the transverse components $p_x^2+p_y^2$ is replaced by $(2n+1)gB$ and the transverse integral
$\int d^2p/(2\pi)^2$ is replaced with the summation $gB/(2\pi)\sum_{n=0}^{n=\infty}$. Then,
after performing the summation, we obtain the factor $(1-\exp(-\frac{2\pi gB)}{gE}))^{-1}$.
In this way we can easily obtain the formula eq(\ref{n}) with $B\neq 0$ 
by replacing corresponding terms
in eq(\ref{b}) with relevant ones.

Using the formula, we can derive the number densities of the gluons
and monopoles in terms of the background gauge fields $\vec{E}$ and $\vec{B}$. 
Only the difference between the scalar
particles and the gluons ( monopoles ) lies in their masses.
The gluons with spin parallel and the monopoles have imaginary masses, 
i.e. $i\sqrt{2gB}$ and $i\mu$, respectively.
( The gluons with spin anti-parallel have the effective mass $\sqrt{2gB}$. )
Then, by replacing the mass in the formula eq(\ref{n}) with the relevant ones, we obtain

\begin{equation}
\label{nn}
n_g=\frac{gEt\times gB}{(2\pi)^2}\frac{\exp\bigl(\frac{\pi gB}{gE}\bigr)+\exp\bigl(\frac{-3\pi gB}{gE}\bigr)}
{1-\exp(-\frac{2\pi gB)}{gE})}, \quad \mbox{and} \quad
n_m=\frac{g_mBt\times g_mE}{(2\pi)^2}\frac{\exp\bigl(\frac{\pi(\mu^2-g_mE)}{g_mB}\bigr)}
{1-\exp(-\frac{2\pi g_mE)}{g_mB})},
\end{equation}   
where the first term with $\exp\bigl(\frac{\pi gB}{gE}\bigr)$ in $n_g$ 
represents the contribution of the gluons with spin parallel ( Nielsen-Olesen unstable modes ) ,
while the second term with $\exp\bigl(\frac{-3\pi gB}{gE}\bigr)$ 
does the one of the gluons with spin anti-parallel. 
These formlas have been explicitly obtained\cite{ita} in canonical formalism, in which
the gluons and monopoles with imaginary masses $i\eta$ become real particles when
the squares of the energies $E^2=p_z^2+(i\eta)^2>0$ are positive 
with large $p_z=\int^{\infty} dt' gE$ or $p_z=\int^{\infty} dt' g_mB$. 
( Particles with imaginary masses $im$ are virtual, but they become real when the square of their energies
$E^2=p^2-m^2$ is positive with large momentum $p>m$. The real particles can be properly treated
in canonical formalism of quantum field theories. )

It is easy to see from the formulas that the gluons are dominant decay products 
when the initial values $Q^2$ of $gB$ and $gE$ are larger than $\mu^2$,
while the monopoles are dominant ones when the initial values of $gB$ and $gE$ are smaller than $\mu^2$.
This is because the gluon production is accelerated by the decrease of $gE$ according to 
the factor $\exp(\pi gB/gE)$ in $n_g$,
while the monopole production is done by the decrease of $gB$ according to the factor $\exp(\pi \mu^2/gB)$ in $n_m$.   
The gluon ( monopole ) production makes the electric field $gE$ ( magnetic field $gB$ ) decrease.

Using the number densities eq(\ref{nn}) and the equation (\ref{2}), 
we find the evolution equations of $gE$ and $gB$,

\begin{equation}
\frac{d(gE)}{d\tau}=-\frac{\alpha_s gE\,gB}{2\pi}\frac{\exp\bigl(\frac{\pi gB}{gE}\bigr)+\exp\bigl(\frac{-3\pi gB}{gE}\bigr)}
{1-\exp(-\frac{2\pi gB)}{gE})} \quad \mbox{and} \quad
\frac{d(gB)}{d\tau}=-\frac{gE\,gB}{2\pi\alpha_s}\frac{\exp\bigl(\frac{\pi(\mu^2-g_mE)}{g_mB}\bigr)}
{1-\exp(-\frac{2\pi g_mE)}{g_mB})}
\end{equation}
with $\tau\equiv t^2$.
We solve the equations with the initial conditions $gE(\tau=0)=gB(\tau=0)=Q^2$.
These are equations governing the production of the gluons and monopoles and the decay of the background gauge fields.
They are very rough approximate formulas for corresponding equations derived in the classical statistical field theories.

\vspace{0.2cm}
Up to now, we derive the evolution equations of the gauge fields in SU(2) gauge theory.
In the case of SU(3) gauge theory, we have three types of the off-diagonal gluons\cite{su3}
and magnetic monopoles.
The gluons are described by the gauge fields,

\begin{equation}
\Phi_1^{\nu}=\frac{A_1^{\nu}+iA_2^{\nu}}{\sqrt{2}}, \quad 
\Phi_2^{\nu}=\frac{A_4^{\nu}+iA_5^{\nu}}{\sqrt{2}}, \quad
\Phi_3^{\nu}=\frac{A_6^{\nu}-iA_7^{\nu}}{\sqrt{2}},
\end{equation}
where the indices $a$ of $A_a^{\nu}$ denote color degrees of freedom.
The gluons couple with the background color electric $E$ and magnetic fields $B$
in maximal Abelian space, 

\begin{equation}
E^i=\delta^{i,z}E\Bigl(\cos(\theta)\lambda_3+\sin(\theta)\lambda_8\Bigr) \quad \mbox{and} \quad
B^i=\delta^{i,z}B\Bigl(\cos(\theta)\lambda_3+\sin(\theta)\lambda_8\Bigr).
\end{equation}
where the angle $\theta$ describes the direction of the gauge fields in the maximal Abelian space 
spanned by the diagonal Gell-Mann matrices $\lambda_3$ and $\lambda_8$.
The angle $\theta$ takes a value in a range $-\pi/6\le \theta\le \pi/6$ owing to 
the Weyl symmetry.
We take an average over the angle
to obtain final results by assuming the uniform distribution in $\theta $.
The coupling constants of the gluons $\Phi_i$ are given by

\begin{equation}
g_1=g\cos\theta,\quad g_2=\frac{g\Bigl(\cos\theta+\sqrt{3}\sin\theta\Bigr)}{2},\quad
\mbox{and} \quad g_3=\frac{g\Bigl(\cos\theta-\sqrt{3}\sin\theta\Bigr)}{2}
\end{equation} 
respectively.
The each gluon couples with the gauge fields $E$ and $B$ with its coupling constant.  

Similarly, the three types of monopoles $\phi_i$ ( $i=1\sim 3$ ) couple with
dual gauge fields $A_a^{d,\nu}$ through covariant derivative, 
$D_{\nu}\phi_i=(\partial_{\nu}+ig_{m,i}A_{\nu}^d)\phi_i$
where the magnetic charges are given by $g_{m,1}=g_m\cos\theta$, 
$g_{m,2}=g_m(\cos\theta+\sqrt{3}\sin\theta )/2$ and 
$g_{m,3}=g_m(\cos\theta-\sqrt{3}\sin\theta )/2$.

Therefore, we add all the contributions of the three types of the gluons and the monopoles
to obtain the number densities $n_g$ and $n_m$ in SU(3) gauge theory,

\begin{eqnarray}
\label{n}
n_g&=&\sum_{i=1\sim 3}\frac{g_iEt\times g_iB}{(2\pi)^2}\frac{\exp\bigl(\frac{\pi gB}{gE}\bigr)+\exp\bigl(\frac{-3\pi gB}{gE}\bigr)}
{1-\exp(-\frac{2\pi gB)}{gE})}=\frac{3}{2}
\frac{gEt\times gB}{(2\pi)^2}\frac{\exp\bigl(\frac{\pi gB}{gE}\bigr)+\exp\bigl(\frac{-3\pi gB}{gE}\bigr)}
{1-\exp(-\frac{2\pi gB)}{gE})} \nonumber \\
n_m&=&\sum_{i\sim 3}\frac{g_{m,i}Bt\times g_{m,i}E}{(2\pi)^2}
\frac{\exp\bigl(\frac{\pi(\mu^2-g_{m,i}E)}{g_{m,i}B}\bigr)}
{1-\exp(-\frac{2\pi g_mE)}{g_mB})}
\end{eqnarray}
where we used the formulae $g_iB/g_iE=gB/gE$, $g_{m,i}E/g_{m,i}B=g_mE/g_mB$ and $\sum_{i=1\sim 3}g_i^2=3g^2/2$.

Therefore, the evolution equations are given by

\begin{eqnarray}
\frac{d(gE)}{d\tau}&=&-\frac{(6\cos\theta-\cos3\theta)}{4}
\frac{\alpha_s gE\,gB}{2\pi}\frac{\exp\bigl(\frac{\pi gB}{gE}\bigr)+\exp\bigl(\frac{-3\pi gB}{gE}\bigr)}
{1-\exp(-\frac{2\pi gB)}{gE})} \\
\frac{d(gB)}{d\tau}&=&-\frac{gE\,gB}{2\pi\alpha_s^2}\biggl(\cos^3\theta
\exp\Bigl(\frac{\pi\mu^2\alpha_s}{gB\cos\theta}\Bigr)+
2^{-3}(\cos\theta+\sqrt{3}\sin\theta)^3\exp\Bigl(\frac{2\pi\mu^2\alpha_s}
{gB(\cos\theta+\sqrt{3}\sin\theta)}\Bigr) \nonumber \\
&+&2^{-3}(\cos\theta-\sqrt{3}\sin\theta)^3 \exp\Bigl(\frac{2\pi\mu^2\alpha_s}
{gB(\cos\theta-\sqrt{3}\sin\theta)}\Bigr)\biggr)
\frac{\exp(-\frac{\pi gE}{gB})}{1-\exp(-\frac{2\pi gE}{gB})}
\end{eqnarray}
with the initial conditions $gE(\tau=0)=gB(\tau=0)=Q^2$.
After solving the equations with $\theta$ fixed, we calculate the number densities $n_g(\theta)$ and
$n_m(\theta)$,
and take the average  
$\bar{n}\equiv \int_{-\pi/6}^{\pi/6} \frac{d\theta}{\pi/3}\, n(\theta)$.

\section{results}
\label{6}

We have solved the equations numerically with the use of the value $\mu=0.5$ GeV and
the running coupling constant $\alpha_s(Q)=g^2/4\pi=\alpha_c(1+\frac{9\alpha_c}{4\pi}\log(Q^2/T_c^2))^{-1}$
with $\alpha_c=0.9$ and $T_c=0.16$GeV used in the reference\cite{gyu}.
We take four values $\theta=0,\pi/12,\pi/9$ and $\pi/6$ and take average of 
the densities $n_g$ and $n_m$ over the values $\theta$.

\vspace{0.1cm}
We show the number density of gluons $n_g$ in Fig.1. We can see that 
the number density per $Q^3$ of the gluons $n_g/Q^3$ is very small when
small $Q<0.35$GeV, but it  
grows as $Q$ increases and approaches an approximate constant
when large $Q>0.5$ GeV.
We also show the number density per $Q^3$ of magnetic monopoles $n_m/Q^3$ in Fig.1.
We find that the monopole production is suppressed when large $Q>0.5$GeV, while the production 
is enhanced when small $Q<0.4$GeV.

\begin{figure}[htb]
 \centering
  \includegraphics*[width=75mm]{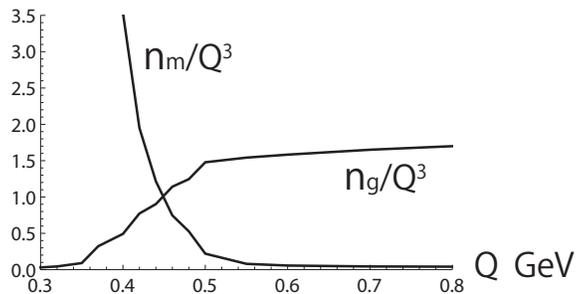}
  \caption{The number density of gluons are suppressed when small $Q<0.4$GeV, while the number density
of monopoles is enhanced.}
     \label{f1}
\end{figure}
\begin{figure}[htb]
 \centering
  \includegraphics*[width=75mm]{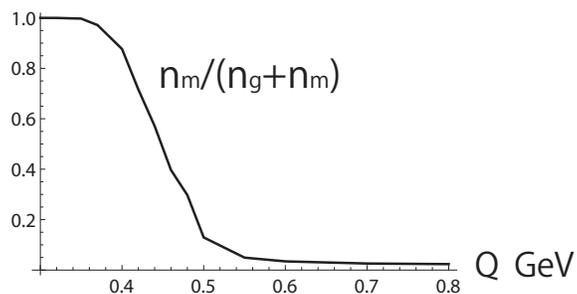}
  \caption{The monopoles are dominant when small $Q<0.4$GeV, while the gluons are dominant when large $Q>0.5$GeV.}
     \label{f2}
\end{figure}
\begin{figure}[htb]
 \centering
  \includegraphics*[width=75mm]{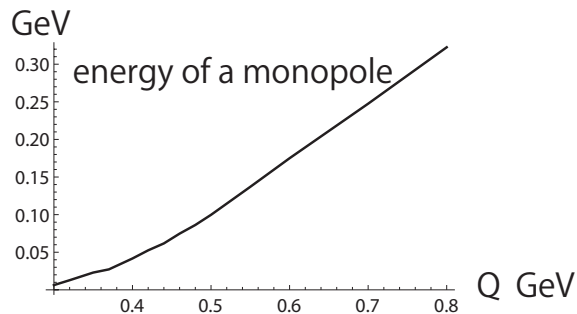}
  \caption{The average kinetic energy of a monopole becomes smaller as $Q$ becomes smaller. 
Using the energy conservation $n_m/Q^3\times (\rm kinetic \,\,energy )\simeq Q/g^2$, we can see how the kinetic energy decreases with $Q$.} 
     \label{f3}
\end{figure}
\begin{figure}[htb]
 \centering
  \includegraphics*[width=75mm]{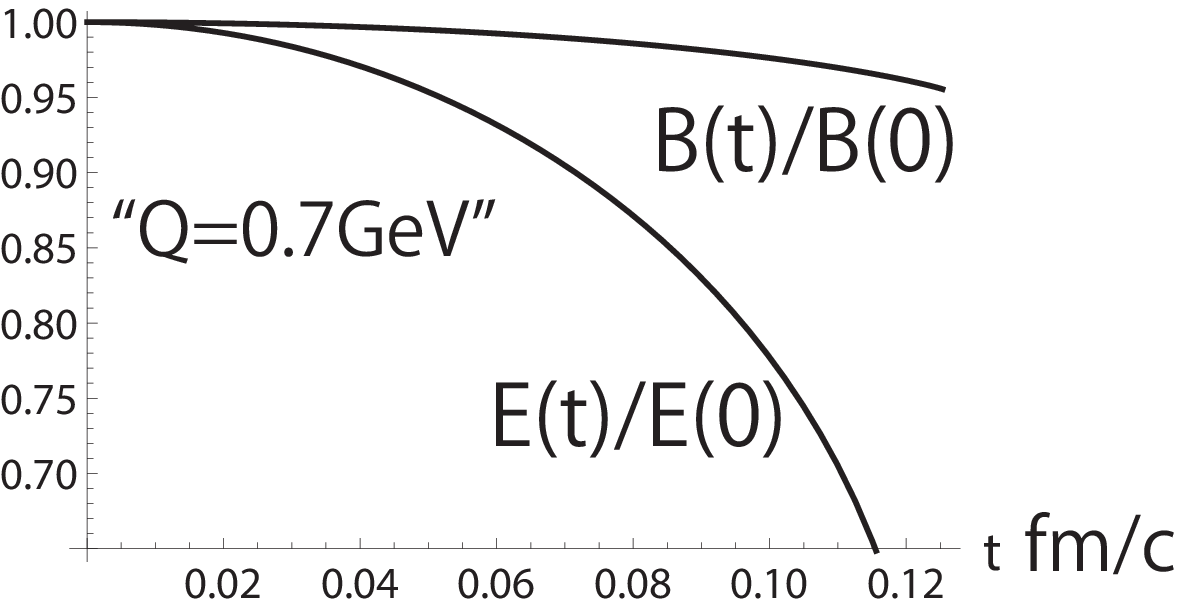}
  \caption{At $Q=0.7$GeV, the electric field fast decays, 
while the magnetic field slowly decays.}
     \label{f4}
\end{figure}
\begin{figure}[htb]
 \centering
  \includegraphics*[width=75mm]{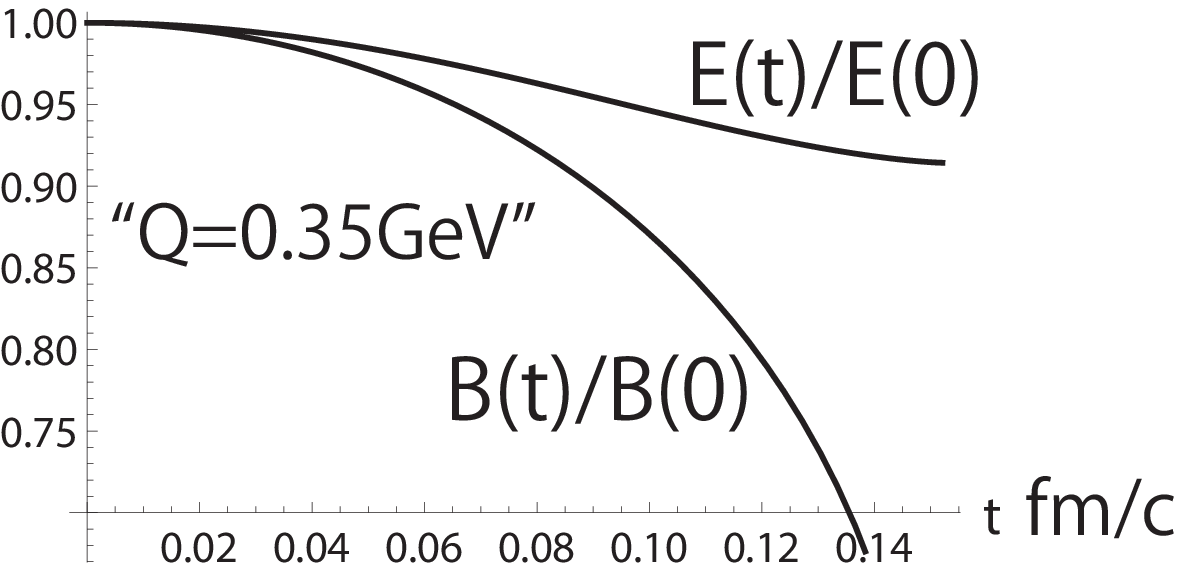}
  \caption{At $Q=0.35$GeV, the electric field slowly decays, while the 
magnetic field fast decays.}
     \label{f5}
\end{figure}
\begin{figure}[htb]
 \centering

  \includegraphics*[width=65mm]{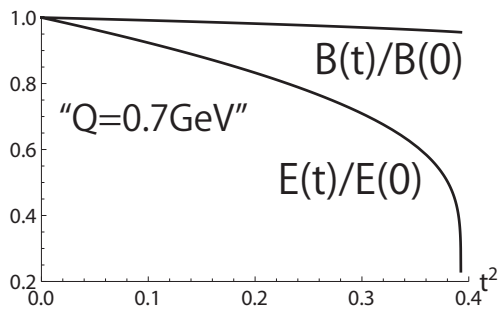}
  \caption{ At $Q=0.7$GeV, we can see that the rapid decay of the electric field starts 
at around $t^2\sim 0.38(\mbox{fm/c})^2$.}
     \label{f6}
\end{figure}
\begin{figure}[htb]
 \centering

  \includegraphics*[width=75mm]{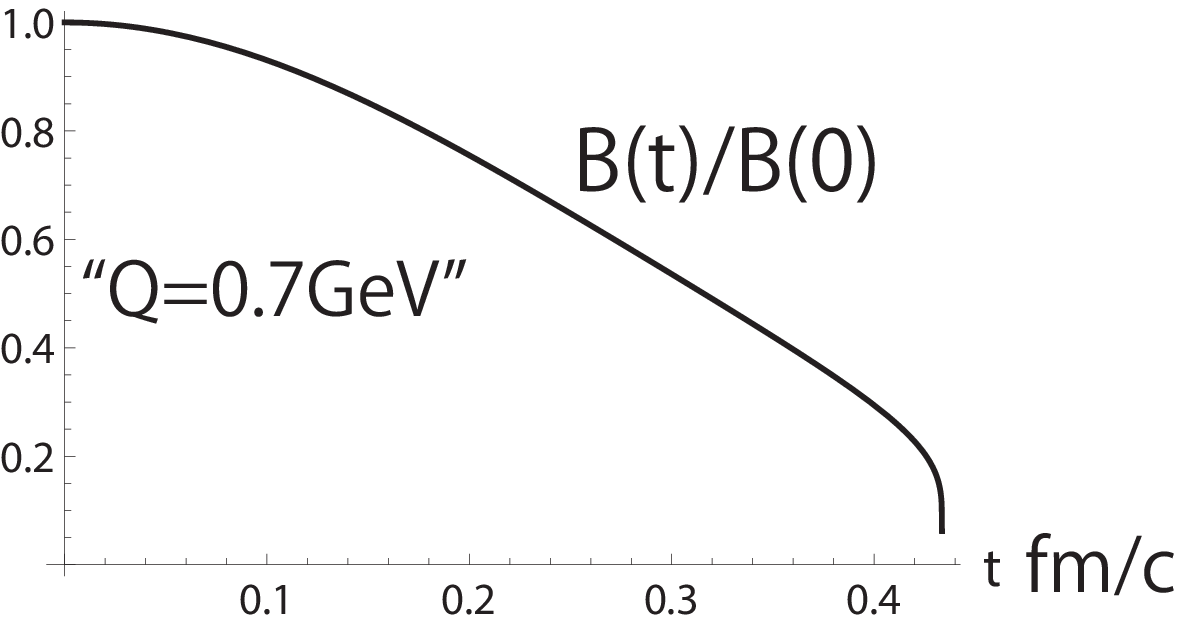}
  \caption{ We depict how the magnetic field slowly decays at $Q=0.7$GeV after the electric field vanishes $E(t\simeq 0.12\rm fm/c)=0$.}
     \label{f7}
\end{figure}

We show the fraction of the monopoles $n_m/(n_g+n_m)$ in Fig.2.
We find that  the gluons are dominant decay products of the strong gauge fields
with $Q>0.5$GeV$\sim 3T_c$,
while the monopoles are dominant ones of the weak gauge fields with $Q<0.4$GeV$\sim 2T_c$.  
The dominance of the gluons ( monopoles ) comes from the factor $\exp(\pi gB/gE)$ in $n_g$ ( $\exp(\pi \mu^2/gB)$ in $n_m$ )
when $gE$ ( $gB$ ) decreases with the production of the gluons ( monopoles ).

\vspace{0.1cm}
These features shown in Fig.1$\sim $Fig.3 are the very similar to those of the QGMPs recently proposed 
\cite{gyu} if we identify $Q$ as temperatures $T$, although the decay products do not interact with each other and are never thermalized in our discussions.
There are no affirmative reasons for the identification $T\sim Q$.
But we would like to point out that 
the energy density of the
gluons and monopoles is given by $Q^4/g^2$, while the energy density of thermalized massless particles
is given by $\pi^2f_{\rm eff}T^4/30$; $f_{\rm eff}$ denotes the number of the species of the massless particles.   
Thus, it is not unreasonable to adopt the identification $Q\simeq gT(f_{\rm eff}\pi^2/30)^{1/4}$.

Using these results, we can see how the average kinetic energies $\epsilon_{g,m}$ of the gluons and the monopoles behave when $Q$ changes.
In order to see it, we note the energy conservation $n_g\times \epsilon_g+n_m\times \epsilon_m\simeq Q^4/g^2$.
When $Q$ is large, the gluons are dominant products and the average kinetic energy $\epsilon_g$ of the gluons behaves such that $\epsilon_g\sim Q$, since
$n_g/Q^3$ is approximately constant for large $Q$. On the other hand, when $Q$ is small, the monopoles are dominant products and 
the kinetic energy $\epsilon_m$ becomes smaller as $Q$ becomes smaller. But we note that $Q$ must be larger than $\Lambda_{QCD}$ or $T_c$.
Thus we can not take the limit $Q\to 0$. Thus, we wish to see how small the kinetic energies of the monopoles are.
For the purpose, 
we use the formulas $n_m$ in eq(\ref{n}) even with small $p_z$, 
although the formulas are only valid 
for the large momentum $p_z >\sqrt{gB}=Q$ (  $ p_z>\mu$ ).
We show in Fig.3 an average kinetic energy of a monopole given  by
$\epsilon_m\equiv |p_z|=\int_0^{t_c} dt n_m(t) g_mB(t)/n_m(t_c)$ where
$t_c$ denotes a time at which
the densities $n_m$ have been evaluated. 
The energy is the one
acquired by a monopole as a result of the acceleration by the magnetic field $B$. 
We find that $\epsilon_m$ is approximately ten times smaller than $Q$ for small $Q$.
As we show below, the small kinetic energy of the monopoles causes large magnetic resistance of the monopole plasmas.

\vspace{0.1cm}
We show in Fig.4 and Fig.5 how $E$ and $B$ decay with time $t$.
When large $Q>0.6$GeV, the electric field fast decays while the magnetic field slowly decays.
On the other hand, when small $Q<0.4$GeV,
the magnetic field fast decays while the electric field slowly decays.
In particular, we should note that the gauge fields smoothly decay in the beginning and then
they start to rapidly decay at a 
time $t_c$ as shown in Fig.6 where 
we use the unit $t^2$ in horizontal axis to represent more clearly how rapid they decay.
The time $t_c$ has been used in the evaluation of $n_g(t=t_c)$ and $n_m(t=t_c)$.

Finally, we show in Fig.7 
how the magnetic field $B$ decays after the electric field $E$ vanishes when $Q=0.7$GeV.
Obviously, the decay proceeds very slowly compared with the decay of the electric field shown in Fig.4.
Similarly we can show that the magnetic fields rapidly decay at first and then the electric fields
slowly decay when $Q=0.35$GeV.
As we have stated before, the slow decays of the remaining gauge fields are not correct because
the models of the gauge fields are not applicable to their decays. Namely,
the decay of the magnetic fields with large $Q$ or small $\alpha_s(Q)\ll1 $
can not be discussed in the model of dual superconductors where the monopoles strongly interact with each other.
Similarly, the decay of the electric fields with small $Q$ or $\alpha_s(Q)\sim 1$ can not be discussed in the perturbative model of the gluons.
We show in the next section that 
the remaining gauge fields rapidly vanish owing to the dissipation of their energies in the background gluon or monopole plasmas
which are produced at first in the Schwinger mechanism. 
Thus, their decays do not change the result that the gluons are dominant decay products of the weakly coupled glasmas and the monopoles are dominant ones
in the strongly coupled glasmas.

\vspace{0.1cm}
We have used the parameter $\mu=0.5$GeV in the calculation.
Physical quantities such as $n_g$ or $n_m$ depend on
$\mu$. When we use different values $\mu$,
the whole behaviors of $n_g/Q^3$ and $n_m/Q^3$ in $Q$ do not change. But
the point $Q_c(\mu)$ at which $n_g(Q_c)/Q_c^3$ is equal to $n_m(Q_c)/Q_c^3$ is different.
For example, when $\mu$ becomes larger than $\mu=0.5$GeV, 
$Q_c(\mu)$ becomes larger than $Q_c(\mu=0.5\rm GeV)$.
Namely,
the monopole dominance over the gluons arises at larger $Q$ than $Q_c\simeq 0.45$GeV shown in Fig.2.
The large imaginary mass of the monopoles enhances 
the spontaneous production of 
the monopoles owing to the factor $\exp(\pi\mu^2/gB)$ in $n_m$.
 
In our previous paper\cite{iwazaki} 
we have discussed the decay of the gauge fields
based on the classical statistical field theory,  
where we have used the values $\mu=0.7$GeV and $Q=0.34$GeV. We 
have found that the magnetic fields vanish in a very short time $<0.1$fm/c, while
the electric fields decay very slowly. These results are consistent with
the present analysis.

\section{discussion and conclusion}
\label{7}

We have shown that
when $Q$ is large such as $Q>0.5$GeV, the color electric fields 
rapidly decay into the gluons at first and then, the remaining color magnetic fields slowly decay into the monopoles. 
But, the decay mechanism of the magnetic fields is not reliable for strong magnetic fields with large $Q$ ( in other words, large $g_m$ ).
Here, we would like to discuss how the remaining magnetic fields rapidly decay without the monopole production. As we show below,
the gluon plasmas produced by the decay of the electric fields have color electrical conductivities $\sigma$ proportional to $Q$ for large $Q$.
Then, the magnetic fields $B$ vanish in the plasmas within a time of the order of $Q^{-1}$ according to
the Ampere's law $\partial_t B=\partial_x^2 B/\sigma$, since they typically possess the momenta $Q$ in reality, $\partial_x^2 B\sim -Q^2B$.
( Although we have assumed the spatial homogeneity of $B$, it typically has momenta of the order of $Q$ or $Q_s$. ) 
Therefore, the magnetic fields rapidly decay for large $Q$ without the monopole production after the electric fields decay.
The fact does not change our result that the gluons are dominant decay products of the weakly coupled glasmas with large $Q$.

We explain that the conductivities are of the order of $Q$ for large $Q$. The conductivities are roughly given by $g^2n_gt_f/\epsilon$ where
$\epsilon$ denotes an average kinetic energy of a gluon and $t_f$ does mean free time of gluons.
$t_f$ is defined by $l/v$ with mean free path $l$ and velocity $v=1$ of gluons. On the other hand, 
the mean free path $l$ is obtained in terms of the collision cross section
$S=\pi r^2$ of the gluons such as $l=1/(n_gS)$ where $r$ is determined by equating the potential energy $g^2/r$ with the kinetic energy $\epsilon_g$.
It is easy to see that $S$ is of the order of $Q^{-2}$ and $t_f=l \sim Q^{-1}$ since $\epsilon_g \sim Q$ and $n_g\sim Q^3$ for large $Q$ as we have shown in Fig.1.
( Here we note the energy conservation $n_g\times \epsilon_g\simeq Q^4/g^2$. )
Therefore, we find that $\sigma$ is of the order of $Q$ for large $Q$. 

 \vspace{0.1cm} 
On the other hand, we have shown that when $Q$ is small but larger than $\Lambda_{QCD}$, the magnetic fields rapidly decay into the monopoles at first and then, 
the electric fields slowly decay into the gluons. But the decay mechanism of the electric fields is not reliable for the weak electric fields with small $Q$ 
( in other words, large $g$ ). We would like to discuss that the electric fields rapidly vanish without the gluon production. They would decay
owing to the large magnetic resistance of the monopole plasmas, in other words, small magnetic conductance $\sigma_m$.
Namely, the monopoles produced by the decay of the magnetic fields have smaller kinetic energies as $Q$ becomes smaller.
Then the collision cross section $S=\pi r^2$ among the monopoles becomes larger as $Q$ becomes smaller. 
That is, the cross section is determined by solving the equation such as the potential energy $g_m^2/r$ equal to the kinetic energy $\epsilon_m$.
Thus, $S=\pi (g_m^2/\epsilon_m )^2$. This implies that the monopole plasmas have
small magnetic conductance for small $Q$. Actually,
the magnetic conductance $\sigma_m$ is given by  $g_m^2n_mt_f/\epsilon_m\simeq g_m^2/(\epsilon_m v S)\simeq \sqrt{2\epsilon_m M}/(2\pi g_m^2)$
where we assume $\epsilon_m= Mv^2/2$ with the mass $M$ and velocity $v$ of the monopoles. 
Although the mass of the monopoles is imaginary when they are produced, the monopoles would acquire real mass $M$ after their production.
Thus,
the decay time $\tau\sim Q^{-2}\sigma_m$ of the electric fields is approximately estimated such that $\tau \simeq 0.04$fm/c 
when $\epsilon_m=0.05$GeV, $Q=0.4$GeV, $M=0.4$GeV and $g_m=1$. 
Although the mass $M$ of the monopoles after their productions is unknown and the estimation is rough, 
our result indicates that the electric fields rapidly decay in the 
monopole plasmas.
Therefore, the monopoles are dominant decay products of the weakly coupled glasmas with small $Q$, since the decay of the electric fields does not produce the gluons.

\vspace{0.1cm}
We have shown that the dominant decay products are gluons when large $Q>3T_c=3\times 0.16$GeV, 
while they are monopoles when small $ Q\simeq (1\sim 2)T_c$. These dominant decay products remain the main components 
after their thermalization as proposed in the model of QGMPs. 
The dominant decay products is determined by the comparison between the production rate
$\propto \exp(gB/gE)$ in eq(\ref{nn}) of Nielsen-Olesen unstable modes and the rate  
$\propto\exp((\mu^2-g_mE)/g_mB)$ in eq(\ref{nn}) of monopoles with imaginary mass.
When the initial values $gE=gB=Q^2$ are larger than $(\mu^2-g_mE)$, the Nielsen-Olesen unstable modes are dominantly produced in the initial
stage of the production. Then, $gE$ decreases faster than $gB$, which accelerates the production of the unstable modes. Thus, 
the dominant products are gluons when large $Q$. On the other hand, when the initial values $gE=gB=Q^2$ are smaller than $(\mu^2-g_mE)$,
the monopoles are dominantly produced. Then, $gB$ decreases faster than $gE$, which accelerates the production of the monopoles.
This is our production mechanism of the dominant particles.

\vspace{0.1cm}
Although we have not quantitatively discussed the momentum distribution of the prethermalized gluons and monopoles,
we can qualitatively discuss the dominance of
the soft gluons with $p^2=p_z^2+p_T^2\ll Q^2$ produced in the decay of the glasmas with large $Q$. 
This is because the Nielsen-Olesen modes with smaller longitudinal momentum $p_z^2$, which grow as $\exp(t\sqrt{gB-p_z^2})$, are produced
more abundantly and their typical transverse momentum $p_T^2\sim gB$ vanish as $gB$ vanishes with the dissipation in the gluon gas.
We should remember that the fields of the modes are given such that $\Phi_{NO}\propto \exp(-x_t^2gB/4)$ with the transverse coordinates $x_t$. 
Therefore, we find that the gluons are mainly composed of soft modes after the decay of the glasmas.
Similarly, the soft modes of the monopoles are dominantly produced in the decay of the glasmas with small $Q$.
( As a result, the monopole condensation may arise since the soft modes with almost zero momentum are mainly produced 
in the limit $Q\to T_c$ as numerically shown in eq(\ref{6}).
This leads to the quark confinement. )
The result of the soft gluon production is consistent with the previous studies\cite{dis} using classical statistical lattice simulations.

\vspace{0.1cm}
It has recently discussed\cite{cs} by using classical statistical simulations that the topological transition associated with the Chern-Simons number $N_{cs}$
is enhanced in the early stage of the weakly coupled glasma evolution. That is, the number $N_{cs}$ rapidly increases ( or decreases ) in the stage.
Since $dN_{cs}/dt$ is proportional to $\int d^3x \vec{E}\cdot\vec{B}$, it is easy to see in our analysis that the rapid decay of the electric fields
leads to the rapid change of the number $N_{cs}$. Similarly,
we can show that the rapid topological transition may arise in the early stage of the strongly coupled
glasma evolution in which the magnetic fields rapidly decay.
In this way we can understand the result of the elaborate numerical simulations\cite{cs} simply by 
using the Schwinger mechanism.

\vspace{0.1cm}
Using the model of the dual superconductors of the monopoles,
we have shown that the gluons are dominant decay products of the weakly coupled glasmas with large $Q$,
while the monopoles are dominant ones of the strongly coupled glasmas with small $Q$.
Although our evolution equations of $n_g$ and $n_m$ is very rough,
the roles of the monopoles in the strongly coupled QCD physics is clarified.
Our results support the significance\cite{gyu,mono} of the monopoles in the strongly coupled QGPs.
More rigorous treatment of the monopoles in these strongly coupled QCD physics with low temperatures $T>T_c$ or small saturation momenta $Q_s>\Lambda_{QCD}$ 
is needed to confirm their roles mentioned above.

\vspace{0.2cm}
The author
expresses thanks to the members of KEK
for their useful discussions.



\end{document}